\documentclass[superscriptaddress,aps,9pt,twocolumn]{revtex4-1}
\usepackage[utf8]{inputenc}
\usepackage{graphicx}
\usepackage{verbatim}
\usepackage{amsmath}
\usepackage{amssymb}
\usepackage{bm}
\usepackage{braket}

\begin{document}

\title{Towards Dynamic Simulations of Materials on Quantum Computers}

\author{Lindsay Bassman}
\affiliation{Collaboratory for Advanced Computing and Simulations, University of Southern California, Los Angeles, CA 90089, United States of America}
\author{Kuang Liu}
\affiliation{Collaboratory for Advanced Computing and Simulations, University of Southern California, Los Angeles, CA 90089, United States of America}
\author{Aravind Krishnamoorthy}
\affiliation{Collaboratory for Advanced Computing and Simulations, University of Southern California, Los Angeles, CA 90089, United States of America}
\author{Thomas Linker}
\affiliation{Collaboratory for Advanced Computing and Simulations, University of Southern California, Los Angeles, CA 90089, United States of America}
\author{Yifan Geng}
\affiliation{Collaboratory for Advanced Computing and Simulations, University of Southern California, Los Angeles, CA 90089, United States of America}
\author{Daniel Shebib}
\affiliation{Collaboratory for Advanced Computing and Simulations, University of Southern California, Los Angeles, CA 90089, United States of America}
\author{Shogo Fukushima}
\affiliation{Department of Physics, Kumamoto University, Kumamoto 860-8555, Japan}
\author{Fuyuki Shimojo}
\affiliation{Department of Physics, Kumamoto University, Kumamoto 860-8555, Japan}
\author{Rajiv K. Kalia}
\affiliation{Collaboratory for Advanced Computing and Simulations, University of Southern California, Los Angeles, CA 90089, United States of America}
\author{Aiichiro Nakano}
\affiliation{Collaboratory for Advanced Computing and Simulations, University of Southern California, Los Angeles, CA 90089, United States of America}
\author{Priya Vashishta}
\affiliation{Collaboratory for Advanced Computing and Simulations, University of Southern California, Los Angeles, CA 90089, United States of America}

\begin{abstract}
A highly anticipated application for quantum computers is as a universal simulator of quantum many-body systems, as was conjectured by Richard Feynman in the 1980s. The last decade has witnessed the growing success of quantum computing for simulating static properties of quantum systems, i.e., the ground state energy of small molecules. However, it remains a challenge to simulate quantum many-body dynamics on current-to-near-future noisy intermediate-scale quantum computers.  Here, we demonstrate successful simulation of nontrivial quantum dynamics on IBM’s Q16 Melbourne quantum processor and Rigetti’s Aspen quantum processor; namely, ultrafast control of emergent magnetism by THz radiation in an atomically-thin two-dimensional material.  The full code and step-by-step tutorials for performing such simulations are included to lower the barrier to access for future research on these two quantum computers. As such, this work lays a foundation for the promising study of a wide variety of quantum dynamics on near-future quantum computers, including dynamic localization of Floquet states and topological protection of qubits in noisy environments.
\end{abstract}

\maketitle
\section{Introduction}
Quantum computers perform computation with two-state, quantum-mechanical systems that serve as quantum bits, or qubits.  Qubits can take advantage of purely quantum mechanical properties, such as superposition and entanglement, to outperform the most advanced classical supercomputers for certain classes of problems.  Quantum computers that are currently and soon-to-be available, dubbed Noisy Intermediate-Scale Quantum (NISQ) \cite{preskill2018quantum} computers, comprised of $\mathcal{O}(10^0-10^2)$ qubits, are beginning to be put to use in scientific research, showing great potential as universal computers for simulating quantum many-body systems, an idea first conceived by Richard Feynman \cite{feynman1999simulating} in the early 1980s and later elaborated by Seth Lloyd and his collaborator \cite{lloyd1996universal,abrams1997simulations}. 

Pioneering work showed that simulating systems of fermionic particles on quantum computers could be carried out with polynomial complexity \cite{ortiz2001quantum,zalka1998simulating} (as opposed to the exponential complexity on classical computers).  Subsequent work demonstrated how chemical properties of such systems can be gleaned from the quantum computer \cite{aspuru2005simulated,lidar1999calculating}, culminating in the first experimental implementation of such a simulation on a quantum computer in 2010 \cite{lanyon2010towards}.  The vast majority of simulations carried out on quantum computers since then have been static calculations \cite{colless2018computation,kandala2017hardware,omalley2016scalable,Du2010nmr,hempel2018quantum} with time-independent Hamiltonians.  However, much stands to be learned from the dynamic simulation of systems governed by time-dependent Hamiltonians, though studies of this kind are still in their infancy \cite{lamm2018simulation,wiebe2011simulating}.

One class of promising quantum many-body dynamics problems to be addressed on NISQ computers is the ultrafast control of emergent quantum-mechanical properties by electromagnetic radiation in atomically-thin layered materials (LMs).  Functional LMs will dominate materials science in this century \cite{geim2013van}.  The attractiveness of LMs lies not only in their outstanding electronic, optical, magnetic and chemical properties, but also in the possibility of easily tuning these properties on demand within picoseconds by an external stimulus like electromagnetic radiation \cite{basov2017towards,lin2017ultrafast,Tung2019anisotropic}.  Especially promising is using terahertz (THz) radiation to directly excite specific phonon modes in the material, which in turn modify atomic arrangement, amounting to ultrafast control of electronic properties in the material \cite{li2019terahertz}.  

Of particular interest is controlling magnetism in LMs \cite{kochat2017re,shin2018phonon}.  In a recent experimental study, small levels of emergent magnetism were observed in single-layer, Re-doped MoSe\textsubscript{2}, a prototypical LM comprised of all nonmagnetic elements \cite{kochat2017re}.  Separately, a theoretical study predicted net magnetization in a similar nonmagnetic LM created by the controlled excitation of a specific phonon mode \cite{shin2018phonon}.  Since THz photoexcitation of such LMs has been shown to excite specific phonon modes on sub-picosecond time scales \cite{lin2017ultrafast,bassman2018electronic}, in principle, it could be used to control magnetism in LMs on ultrafast timescales.

The study of such dynamic magnetism in LMs is inherently a quantum many-body problem onto which near-future NISQ computers may be able to provide valuable insights.  As an early proof-of-concept, we use IBM’s Q16 Melbourne quantum computer and Rigetti’s Aspen quantum computer to simulate a simplified model of Re-doped, monolayer MoSe\textsubscript{2} with a specific phonon mode excited, and measure the average magnetization as a function of time.  In an attempt to lower the barrier to entry for future researchers intending to perform quantum dynamics simulations on either quantum computer, we have included all code used for our simulations, as well as step-by-step tutorials for designing quantum circuits, connecting to the quantum processors, and post-processing results for both machines (see Supplemental Material).  Our work provides a compelling proof-of-concept for the accurate simulation of a real material under a time-dependent Hamiltonian on a quantum computer.  Furthermore, the general framework we present for performing such dynamic material simulations can easily be extended to similar or larger material systems with the adjustment of a few parameters.  

\section{Theory}
\subsection{Theoretical Model}
\begin{figure*}
	\centering
	\includegraphics[scale=0.7]{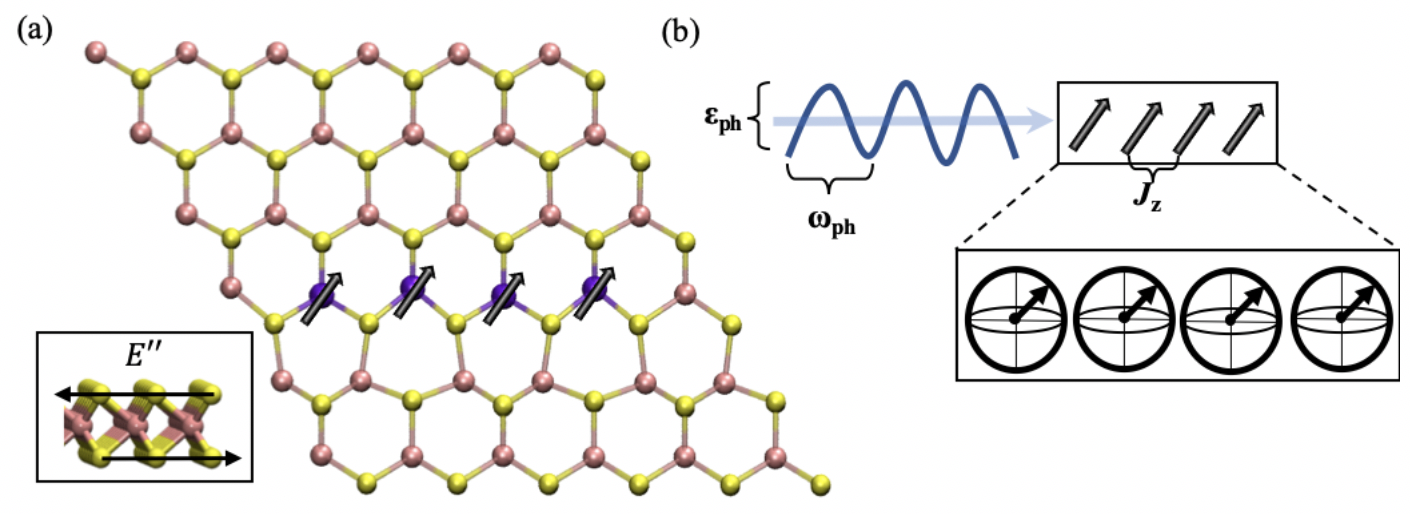}
	\caption{Schematic showing how a simplified model of Re-doped monolayer MoSe\textsubscript{2} is mapped onto the qubits of a quantum computer.  (a) Top-down view of the Re-doped MoSe\textsubscript{2} monolayer where Mo, Se, and Re atoms are depicted by pink, yellow, and purple spheres, respectively.  Grey arrows are superposed on the Re atoms, representing the spin of the extra, unpaired electron each Re atom possesses.  Inset shows a side view of the material and a representation of the excited $E''$ phonon mode.  (b) Spins of the extra, unpaired electron of each Re atom (grey arrows), with exchange interactions between neighboring spins of strength $J_z$, in the presence of an external magnetic field with frequency $\omega_{ph}$ and amplitude $\epsilon_{ph}$ are mapped onto the qubits of a quantum computer, shown in their Bloch sphere representation.}
	\label{fig:model}
\end{figure*}

In Re-doped MoSe\textsubscript{2} monolayer, Re atoms substituting Mo atoms have been experimentally shown to form clusters \cite{kochat2017re}.  Therefore, we consider a 1D cluster of Re atoms in the monolayer as a simple yet nontrivial quantum many-body testbed, which is amenable for study on currently available NISQ computers.  A schematic of the material with a four-Re-atom cluster is shown from the top view in Figure \ref{fig:model}a.  Since each Re atom has one additional, unpaired electron compared to the Mo atom it replaces, we map the spin of each of those electrons to the spin of a qubit on the quantum computer, as depicted in Figure \ref{fig:model}b.  We describe the magnetism using the Ising model, where the exchange interaction strength $J_z$ is computed from first principles (see Supplemental Material for details of calculation).  We simulate excitation of the $E''$ phonon mode in monolayer MoSe\textsubscript{2} (shown in the inset of Figure \ref{fig:model}a), as it has been shown to be the only mode that appreciably couples to the spin motion, i.e. affects the magnetism in the monolayer \cite{shin2018phonon}.  The $E''$ phonon mode gives rise to an effective magnetic field through spin-orbit coupling, which can be incorporated into the Ising model as an oscillatory transverse magnetic field with the $E''$-phonon frequency of $f_{ph} = \frac{\omega_{ph}}{2\pi} =$ 4.8 THz \cite{shin2018phonon}.  The resulting time-dependent Hamiltonian is given by
\begin{equation} \label{hamiltonian}
    H(t) = -J_{z}\sum_{i=1}^{N-1} \sigma_{i}^{z}\sigma_{i+1}^{z} - \epsilon_{ph}cos(\omega_{ph}t) \sum_{i=1}^{N} \sigma_{i}^{x}
\end{equation}
where $\sigma_{i}^{\alpha}$ is the $\alpha$-Pauli matrix acting on qubit $i$, and $\epsilon_{ph}$ is the amplitude of the effective magnetic field produced by the excited $E''$ phonon mode, which is controlled by the fluence of incident electromagnetic radiation in the THz range.  In our simulations, all parameters of Hamiltonian \ref{hamiltonian} are held fixed, except for $\epsilon_{ph}$, which we vary over physically reasonable strengths compared to the coupling strength $J_z$.  For details on how controlled excitation of the $E''$ phonon mode can be used to vary the strength of the effective transverse magnetic field (i.e. the value of $\epsilon_{ph}$), see Supplemental Material.

\subsection{Time Evolution}
We simulate time evolution of our model under the time-dependent Hamiltonian \ref{hamiltonian} by acting on the qubits the time-ordered exponential form of the unitary operator $U(t)\equiv U(0,t)=\mathcal{T} exp(-i\int_{0}^{t}H(t)dt)$ in the atomic unit.  In order to map this unitary operator into a set of gates in a quantum circuit, we first discretize time with a small time step of $\Delta$t, during which $H(t)$ can be regarded as constant \cite{poulin2011quantum}.  We then apply Trotter decomposition \cite{trotter1959on} by splitting the Hamiltonian into components that are each easily diagonalizable on their own: $H(t)=H_x (t)+ H_z$, where $H_x (t)=-J_{X}(t)\sum_{i=1}^{N} \sigma_{i}^{x}$ and $H_z = -J_z \sum_{i=1}^{N-1} \sigma_{i}^{z} \sigma_{i+1}^{z}$.  Thus, the time evolution operator is approximated as

\begin{equation} \label{unitary}
 U(n\Delta t) = \sum_{j=1}^{n-1} e^{-iH_{x}((j+\frac{1}{2})\Delta t)\Delta t} e^{iH_{z}\Delta t} + \mathcal{O}(\Delta t)  
\end{equation}

We note that Hamiltonian \ref{hamiltonian} is in the form of a 1D Ising model with an oscillating, transverse magnetic field.  In the case of a static magnetic field, this model was solved exactly by Pfeuty \cite{pfeuty1970one} following the methods of Lieb, Shultz and Mattis \cite{lieb1961two}.  More recently, Ref. \cite{verstraete2009quantum} proposed applying this method to quantum simulations on quantum computers to efficiently create strongly correlated quantum states and simulate their dynamical evolution for arbitrary times, while Ref. \cite{cervera2018exact} carried out this proposal on IBM’s quantum computer.  While these transformations only apply to strictly 1D chains, it is useful to be able to simulate the spin dynamics of arbitrary clusters, for which no such transformation is known. We have thus employed a more general solution method, Eq. \ref{unitary}, which applies to arbitrary clusters.

\subsection{Quantum Simulation on Quantum Computers}
The basic quantum circuit for simulating the dynamics of the system involves (i) initialization of the qubits, (ii) application of the time-evolution operator to the qubits, and (iii) measurement of each qubit in the computational basis, which is assumed to be the $z$-basis.  Initializing qubits to a desired initial state is a non-trivial task, for which a number of different methods have been proposed \cite{ortiz2001quantum,zalka1998simulating,aspuru2005simulated,mosca2001quantum,soklakov2006efficient,kitaev2008wavefunction,wang2009efficient,Veis2014adiabatic,tubman2018postponing}.  We use a ferromagnetic configuration (all spins up) as our initial state, since it is the ground state of the Hamiltonian (1) in the absence of THz radiation.  The goal of the simulations is to study the dynamics of magnetism by switching on the THz field at time t = 0.  Fortunately, all qubits are initialized in the spin-up position by default on both IBM’s and Rigetti’s quantum processors, and therefore qubit initialization is trivial, i.e. no quantum gates are required.  After qubit initialization, the time-evolution operator given in Eq. \ref{unitary} is translated into a set of quantum gates and applied to the qubits, followed by measurement of all qubits.  In our case, the measurement of interest is the time-dependent, average magnetization of the $N$ qubits along the $z$-direction, given by $\langle m_z(t)\rangle \equiv  \frac{1}{N} \sum_i \sigma_i^z(t)$.  An illustration of a sample quantum circuit simulating evolution to the first time-step for a three-qubit system is included in the Supplemental Material.
Since measurement of the qubits destroys their quantum state, in order to simulate dynamic evolution of the qubits through time, the qubits must always be initialized to their time $t=0$ values before applying the appropriate time-evolution operator $U(n\Delta t)$ (a separate quantum circuit for each time step $n$).  Furthermore, measurement does not give the full quantum state of the qubits; instead each qubit will only return a 0 or 1.  Therefore, the circuit for each time step must be run a large number of times to reconstruct an estimate for the expectation value of the observable. Pseudocode for simulation, as well as the full code, are available in the Supplemental Material.

\section{Results}
\subsection{Theoretical Results}
\begin{figure*}
	\centering
	\includegraphics[scale=1]{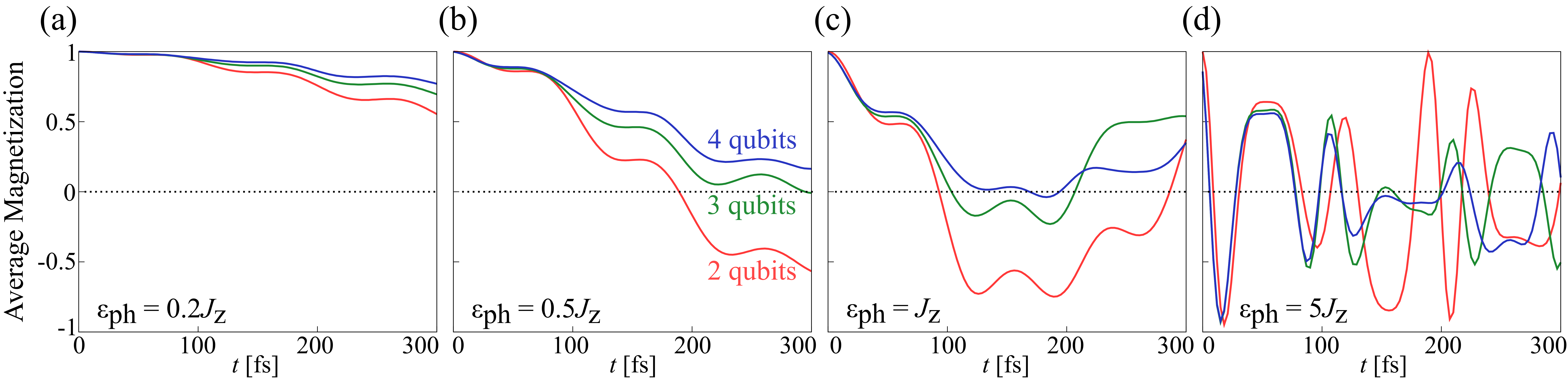}
	\caption{Time evolution of the average magnetization of 2- (red), 3- (green), and 4-qubit (blue) systems with electron-phonon coupling strengths $\epsilon_{ph} = 0.2J_z$ (a),$0.5J_z$ (b), $J_z$ (c) and $5J_z$ (d).  The black dotted line shows zero magnetization.}
	\label{fig:theorectical_results}
\end{figure*}
To establish a ground truth for validation of our quantum computer results, we first calculate the dynamics of our system using a wavefunction simulator (see Supplemental Material for a detailed description).  We perform simulations for 2-, 3-, and 4-qubit systems with various values of $\epsilon_{ph}$, keeping all other values in Hamiltonian \ref{hamiltonian} constant.  A time-step of $\Delta t=3$ fs is used (see Supplemental Material for how this value was chosen).  The initial state of all three systems is defined as all spins up, giving an initial average magnetization of 1.  Since the exchange interaction strength $J_z$ is positive, the Hamiltonian is ferromagnetic and thus, the exchange term (first term) of the Hamiltonian will tend to keep the qubits aligned.  The phonon-induced magnetic field term (second term) of the Hamiltonian, which acts in a direction perpendicular to the initial orientation of the qubits, will tend to push them out of alignment, reducing the average magnetization of the system.  Thus, as we increase the ratio $\frac{\epsilon_{ph}}{J_z}$ , we expect to see larger drops in the average magnetization at the start of the simulations.  

Figure \ref{fig:theorectical_results} shows the resultant time-dependent average magnetization for 2- (red), 3- (green), and 4-qubit (blue) systems, with varying values of $\epsilon_{ph}$, using the wavefunction simulator.  Indeed, we see that as the strength of the electron-phonon coupling constant, $\epsilon_{ph}$, is increased from $0.2J_z$ (Fig. 2a), to $0.5J_z$ (Fig. 2b), to $J_z$ (Fig. 2c) the average magnetization decreases by greater amounts.  

Another notable observation is a size effect across all values of $\epsilon_{ph}$, in which the more qubits the system has, the smaller the phonon-induced change in average magnetization.  We can attribute this to the fact that as the number of qubits in the system increases, the ratio of bulk qubits to edge qubits also increases.  In the simulated finite 1D cluster, edge qubits (those which only have one nearest neighbor) only contribute one exchange interaction term to the Hamiltonian, while center qubits (those which have two nearest neighbors) contribute two exchange terms.  As the ratio of bulk qubits to edge qubits increases, the ratio of exchange interaction terms to transverse magnetic field terms in Hamiltonian also increases.  Therefore, increasing numbers of qubits reduce the effects of the transverse magnetic field, and thus the average magnetization is reduced by smaller amounts in systems with more qubits.  When the magnetic field becomes much stronger than the coupling strength, $5J_z$ (Fig. 2d), the qubits are initially all rapidly flipped, but then proceed to cycle at different rates depending on how many qubits are present in the system.

\subsection{Experimental Results}
\begin{figure*}
	\centering
	\includegraphics[scale=1]{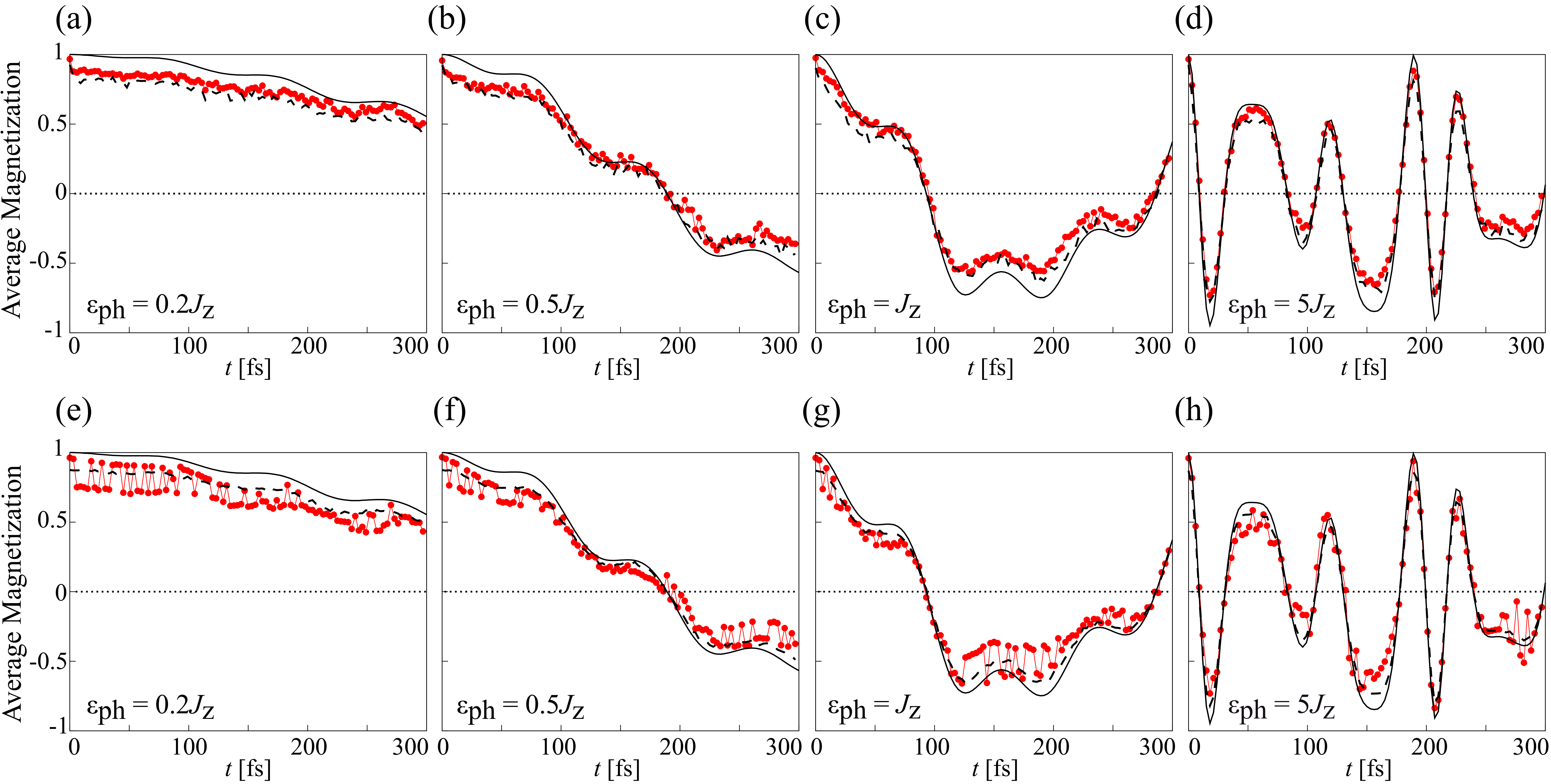}
	\caption{Simulation results for a 2-qubits system (red dots) on the IBM quantum processor (a-d) and the Rigetti Aspen quantum processor (e-h), compared to theoretical results from simulated noisy qubits (black, dashed lines) and the wavefunction simulator (black, solid lines).  The black dotted lines show zero average magnetization.  Results are shown for varying electron-phonon coupling strengths $\epsilon_{ph} = 0.2J_z$ (a,e), $0.5J_z$ (b,f), $J_z$ (c,g) and $5J_z$ (d,h).}
	\label{fig:experimental_results}
\end{figure*}
We perform the same simulations on IBM’s Q16 Melbourne quantum processor and Rigetti’s Aspen quantum processor.  Details of the two quantum processors can be found in the Supplemental Material.  Figure \ref{fig:experimental_results} shows the results for a 2-qubit system with varying ratios of strengths between the exchange interaction term and the transverse magnetic field term, on the IBM (Fig. \ref{fig:experimental_results}a-d) and Rigetti (Fig. \ref{fig:experimental_results}e-h) quantum computers.  The entire simulation (all time-steps with 1,024 (IBM) or 10,000 (Rigetti) trials per time-step) was run five independent times on each quantum processor, with the average results shown in red circles (a line is included between red circles to guide the eye).  

The results from the quantum computers are compared with those from simulated noisy qubits (black, dashed lines) as well as the results from the wavefunction simulator (black, solid lines). See Supplemental Material for a detailed description of the simulated noisy qubits.
The first thing to note in Fig. \ref{fig:experimental_results} is that while the qubits all start in the spin-up position, the average magnetization at time t=0 is not measured to be 1, as should be expected.  This can be attributed to readout noise, in which qubits that are in the state ‘0’ have a small probability of being measured to be in the ‘1’ state, and vice versa.  Readout noise will plague results in a systematic way throughout the entire simulation and is one of the sources of error that can be included in noise models for simulated qubits.

Looking at results across the rest of the simulation time, Fig. \ref{fig:experimental_results} shows good correspondence between those from the quantum processor and the wavefunction simulator, and near overlap of results is found between the quantum processor and simulated noisy qubits.  The close correspondence of these results provides compelling evidence that our current noise models are capturing the largest sources of error on currently available NISQ computers.  

\section{Conclusion}
In this work, we have presented results of dynamic simulations of emergent magnetism in a simplified model of an LM performed on two state-of-the-art quantum computers.  Remarkably good quantitative agreement was found between the results from the quantum computer and those from simulated noisy qubits, and to a slightly lesser extent with those from the wavefunction simulator.  Incorporating error models to account for the decoherence times of the qubits, readout noise, and gate error-rates brought the results from simulated noisy qubits into closer agreement with the quantum computer results, compared to those from the noise-free wavefunction simulator.  This indicates that there is a good understanding of the largest sources of error we currently face on available NISQ computers.  This early proof-of-concept gives hope that near-future NISQ computers, capable of simulating larger systems, may soon be able to deepen understandings of controlled magnetism, as well as other controllable electronic properties in LMs, for use in myriad new technologies.

In the immediate future, NISQ computers with $\mathcal{O}(10^2)$ qubits will allow straightforward extensions of the present work. One possible avenue for future simulation could be the dynamic study of Floquet systems \cite{holthaus1992collapse}.  Here, one could simulate a time-periodic Hamiltonian $H(t)=H_0+H_1(t)$ where $H_1(t)=H_1(t+T)$ harmonically oscillates with a period of $T$, such as the Hamiltonian presented in this work.  The Floquet formalism ]\cite{floquet1883equations} can be used to deal with such time-periodic Hamiltonians, which casts them into effective, quasi-time-independent Hamiltonians, with associated quasi-stationary states, known as Floquet states.  Using the periodic perturbations as a tuning parameter, various Floquet states can be created, leading to novel phases of matter and engineered materials with properties not normally seen in equilibrium.  Powered by quantum computers, dynamic simulations of so called “Floquet engineering” \cite{bukov2015universal}, could help guide experiment to achieve myriad kinds of on-demand functionality in novel metamaterials.  Note that tuning of the $\epsilon_{ph}$ parameter in our simulations could be considered simulated Floquet engineering of the system.  

The second immediate extension of this work could be the study of topological protection of qubits in dissipative environments.  A recent theoretical study suggested the use of nontrivial topological phases associated with edge states in 1D spin-chains for protecting the coherence of qubits in a dissipative environment \cite{venuti2017topological}.  

Finally, there has been a recent surge of interest in antiferromagnetism in LMs \cite{valenzuela2019phase}.  While this paper has focused on emergent ferromagnetism in LMs, a third extension of this work could be the study of antiferromagnetism in these materials.  To demonstrate that our approach is readily applicable to antiferromagnetic cases, Fig. S6 in Supplemental Material shows simulation results from adapting our simulation techniques to antiferomagnetically-coupled chains of spins.  Near-future NISQ computers, with the ability to extend our system, will provide an ideal platform to explore all such possibilities.  

\section*{Funding Information}
This work was supported as part of the Computational Materials Sciences Program funded by the U.S. Department of Energy, Office of Science, Basic Energy Sciences, under Award Number DE-SC0014607.  Quantum computation was performed on the IBM Q16 Melbourne quantum computer and the Rigetti Aspen quantum computer.  The authors would like to thank both IBM and Rigetti for their assistance and access to their resources.

\newpage
\bibliographystyle{apsrev4-1}
\bibliography{Trotter_sim}

\begin{thebibliography}{41}%
\makeatletter
\providecommand \@ifxundefined [1]{%
 \@ifx{#1\undefined}
}%
\providecommand \@ifnum [1]{%
 \ifnum #1\expandafter \@firstoftwo
 \else \expandafter \@secondoftwo
 \fi
}%
\providecommand \@ifx [1]{%
 \ifx #1\expandafter \@firstoftwo
 \else \expandafter \@secondoftwo
 \fi
}%
\providecommand \natexlab [1]{#1}%
\providecommand \enquote  [1]{``#1''}%
\providecommand \bibnamefont  [1]{#1}%
\providecommand \bibfnamefont [1]{#1}%
\providecommand \citenamefont [1]{#1}%
\providecommand \href@noop [0]{\@secondoftwo}%
\providecommand \href [0]{\begingroup \@sanitize@url \@href}%
\providecommand \@href[1]{\@@startlink{#1}\@@href}%
\providecommand \@@href[1]{\endgroup#1\@@endlink}%
\providecommand \@sanitize@url [0]{\catcode `\\12\catcode `\$12\catcode
  `\&12\catcode `\#12\catcode `\^12\catcode `\_12\catcode `\%12\relax}%
\providecommand \@@startlink[1]{}%
\providecommand \@@endlink[0]{}%
\providecommand \url  [0]{\begingroup\@sanitize@url \@url }%
\providecommand \@url [1]{\endgroup\@href {#1}{\urlprefix }}%
\providecommand \urlprefix  [0]{URL }%
\providecommand \Eprint [0]{\href }%
\providecommand \doibase [0]{http://dx.doi.org/}%
\providecommand \selectlanguage [0]{\@gobble}%
\providecommand \bibinfo  [0]{\@secondoftwo}%
\providecommand \bibfield  [0]{\@secondoftwo}%
\providecommand \translation [1]{[#1]}%
\providecommand \BibitemOpen [0]{}%
\providecommand \bibitemStop [0]{}%
\providecommand \bibitemNoStop [0]{.\EOS\space}%
\providecommand \EOS [0]{\spacefactor3000\relax}%
\providecommand \BibitemShut  [1]{\csname bibitem#1\endcsname}%
\let\auto@bib@innerbib\@empty
%</preamble>
\bibitem [{\citenamefont {Preskill}(2018)}]{preskill2018quantum}%
  \BibitemOpen
  \bibfield  {author} {\bibinfo {author} {\bibfnamefont {J.}~\bibnamefont
  {Preskill}},\ }\href {\doibase 10.22331/q-2018-08-06-79} {\bibfield
  {journal} {\bibinfo  {journal} {Quantum}\ }\textbf {\bibinfo {volume} {2}},\
  \bibinfo {pages} {79} (\bibinfo {year} {2018})}\BibitemShut {NoStop}%
\bibitem [{\citenamefont {Feynman}(1999)}]{feynman1999simulating}%
  \BibitemOpen
  \bibfield  {author} {\bibinfo {author} {\bibfnamefont {R.~P.}\ \bibnamefont
  {Feynman}},\ }\href@noop {} {\bibfield  {journal} {\bibinfo  {journal} {Int.
  J. Theor. Phys}\ }\textbf {\bibinfo {volume} {21}} (\bibinfo {year}
  {1999})}\BibitemShut {NoStop}%
\bibitem [{\citenamefont {Lloyd}(1996)}]{lloyd1996universal}%
  \BibitemOpen
  \bibfield  {author} {\bibinfo {author} {\bibfnamefont {S.}~\bibnamefont
  {Lloyd}},\ }\href {\doibase 10.1126/science.273.5278.1073} {\bibfield
  {journal} {\bibinfo  {journal} {Science}\ }\textbf {\bibinfo {volume}
  {273}},\ \bibinfo {pages} {1073} (\bibinfo {year} {1996})}\BibitemShut
  {NoStop}%
\bibitem [{\citenamefont {Abrams}\ and\ \citenamefont
  {Lloyd}(1997)}]{abrams1997simulations}%
  \BibitemOpen
  \bibfield  {author} {\bibinfo {author} {\bibfnamefont {D.~S.}\ \bibnamefont
  {Abrams}}\ and\ \bibinfo {author} {\bibfnamefont {S.}~\bibnamefont {Lloyd}},\
  }\href {\doibase 10.1103/PhysRevLett.79.2586} {\bibfield  {journal} {\bibinfo
   {journal} {Physical Review Letters}\ }\textbf {\bibinfo {volume} {79}},\
  \bibinfo {pages} {2586} (\bibinfo {year} {1997})}\BibitemShut {NoStop}%
\bibitem [{\citenamefont {Ortiz}\ \emph {et~al.}(2001)\citenamefont {Ortiz},
  \citenamefont {Gubernatis}, \citenamefont {Knill},\ and\ \citenamefont
  {Laflamme}}]{ortiz2001quantum}%
  \BibitemOpen
  \bibfield  {author} {\bibinfo {author} {\bibfnamefont {G.}~\bibnamefont
  {Ortiz}}, \bibinfo {author} {\bibfnamefont {J.~E.}\ \bibnamefont
  {Gubernatis}}, \bibinfo {author} {\bibfnamefont {E.}~\bibnamefont {Knill}}, \
  and\ \bibinfo {author} {\bibfnamefont {R.}~\bibnamefont {Laflamme}},\
  }\href@noop {} {\bibfield  {journal} {\bibinfo  {journal} {Physical Review
  A}\ }\textbf {\bibinfo {volume} {64}},\ \bibinfo {pages} {022319} (\bibinfo
  {year} {2001})}\BibitemShut {NoStop}%
\bibitem [{\citenamefont {Zalka}(1998)}]{zalka1998simulating}%
  \BibitemOpen
  \bibfield  {author} {\bibinfo {author} {\bibfnamefont {C.}~\bibnamefont
  {Zalka}},\ }\href@noop {} {\bibfield  {journal} {\bibinfo  {journal}
  {Proceedings of the Royal Society of London. Series A: Mathematical, Physical
  and Engineering Sciences}\ }\textbf {\bibinfo {volume} {454}},\ \bibinfo
  {pages} {313} (\bibinfo {year} {1998})}\BibitemShut {NoStop}%
\bibitem [{\citenamefont {Aspuru-Guzik}\ \emph {et~al.}(2005)\citenamefont
  {Aspuru-Guzik}, \citenamefont {Dutoi}, \citenamefont {Love},\ and\
  \citenamefont {Head-Gordon}}]{aspuru2005simulated}%
  \BibitemOpen
  \bibfield  {author} {\bibinfo {author} {\bibfnamefont {A.}~\bibnamefont
  {Aspuru-Guzik}}, \bibinfo {author} {\bibfnamefont {A.~D.}\ \bibnamefont
  {Dutoi}}, \bibinfo {author} {\bibfnamefont {P.~J.}\ \bibnamefont {Love}}, \
  and\ \bibinfo {author} {\bibfnamefont {M.}~\bibnamefont {Head-Gordon}},\
  }\href {http://science.sciencemag.org/content/sci/309/5741/1704.full.pdf}
  {\bibfield  {journal} {\bibinfo  {journal} {Science}\ }\textbf {\bibinfo
  {volume} {309}},\ \bibinfo {pages} {1704} (\bibinfo {year}
  {2005})}\BibitemShut {NoStop}%
\bibitem [{\citenamefont {Lidar}\ and\ \citenamefont
  {Wang}(1999)}]{lidar1999calculating}%
  \BibitemOpen
  \bibfield  {author} {\bibinfo {author} {\bibfnamefont {D.~A.}\ \bibnamefont
  {Lidar}}\ and\ \bibinfo {author} {\bibfnamefont {H.}~\bibnamefont {Wang}},\
  }\href@noop {} {\bibfield  {journal} {\bibinfo  {journal} {Physical Review
  E}\ }\textbf {\bibinfo {volume} {59}},\ \bibinfo {pages} {2429} (\bibinfo
  {year} {1999})}\BibitemShut {NoStop}%
\bibitem [{\citenamefont {Lanyon}\ \emph {et~al.}(2010)\citenamefont {Lanyon},
  \citenamefont {Whitfield}, \citenamefont {Gillett}, \citenamefont {Goggin},
  \citenamefont {Almeida}, \citenamefont {Kassal}, \citenamefont {Biamonte},
  \citenamefont {Mohseni}, \citenamefont {Powell}, \citenamefont {Barbieri}
  \emph {et~al.}}]{lanyon2010towards}%
  \BibitemOpen
  \bibfield  {author} {\bibinfo {author} {\bibfnamefont {B.~P.}\ \bibnamefont
  {Lanyon}}, \bibinfo {author} {\bibfnamefont {J.~D.}\ \bibnamefont
  {Whitfield}}, \bibinfo {author} {\bibfnamefont {G.~G.}\ \bibnamefont
  {Gillett}}, \bibinfo {author} {\bibfnamefont {M.~E.}\ \bibnamefont {Goggin}},
  \bibinfo {author} {\bibfnamefont {M.~P.}\ \bibnamefont {Almeida}}, \bibinfo
  {author} {\bibfnamefont {I.}~\bibnamefont {Kassal}}, \bibinfo {author}
  {\bibfnamefont {J.~D.}\ \bibnamefont {Biamonte}}, \bibinfo {author}
  {\bibfnamefont {M.}~\bibnamefont {Mohseni}}, \bibinfo {author} {\bibfnamefont
  {B.~J.}\ \bibnamefont {Powell}}, \bibinfo {author} {\bibfnamefont
  {M.}~\bibnamefont {Barbieri}},  \emph {et~al.},\ }\href@noop {} {\bibfield
  {journal} {\bibinfo  {journal} {Nature chemistry}\ }\textbf {\bibinfo
  {volume} {2}},\ \bibinfo {pages} {106} (\bibinfo {year} {2010})}\BibitemShut
  {NoStop}%
\bibitem [{\citenamefont {Colless}\ \emph {et~al.}(2018)\citenamefont
  {Colless}, \citenamefont {Ramasesh}, \citenamefont {Dahlen}, \citenamefont
  {Blok}, \citenamefont {Kimchi-Schwartz}, \citenamefont {McClean},
  \citenamefont {Carter}, \citenamefont {De~Jong},\ and\ \citenamefont
  {Siddiqi}}]{colless2018computation}%
  \BibitemOpen
  \bibfield  {author} {\bibinfo {author} {\bibfnamefont {J.~I.}\ \bibnamefont
  {Colless}}, \bibinfo {author} {\bibfnamefont {V.~V.}\ \bibnamefont
  {Ramasesh}}, \bibinfo {author} {\bibfnamefont {D.}~\bibnamefont {Dahlen}},
  \bibinfo {author} {\bibfnamefont {M.~S.}\ \bibnamefont {Blok}}, \bibinfo
  {author} {\bibfnamefont {M.}~\bibnamefont {Kimchi-Schwartz}}, \bibinfo
  {author} {\bibfnamefont {J.}~\bibnamefont {McClean}}, \bibinfo {author}
  {\bibfnamefont {J.}~\bibnamefont {Carter}}, \bibinfo {author} {\bibfnamefont
  {W.}~\bibnamefont {De~Jong}}, \ and\ \bibinfo {author} {\bibfnamefont
  {I.}~\bibnamefont {Siddiqi}},\ }\href@noop {} {\bibfield  {journal} {\bibinfo
   {journal} {Physical Review X}\ }\textbf {\bibinfo {volume} {8}},\ \bibinfo
  {pages} {011021} (\bibinfo {year} {2018})}\BibitemShut {NoStop}%
\bibitem [{\citenamefont {Kandala}\ \emph {et~al.}(2017)\citenamefont
  {Kandala}, \citenamefont {Mezzacapo}, \citenamefont {Temme}, \citenamefont
  {Takita}, \citenamefont {Brink}, \citenamefont {Chow},\ and\ \citenamefont
  {Gambetta}}]{kandala2017hardware}%
  \BibitemOpen
  \bibfield  {author} {\bibinfo {author} {\bibfnamefont {A.}~\bibnamefont
  {Kandala}}, \bibinfo {author} {\bibfnamefont {A.}~\bibnamefont {Mezzacapo}},
  \bibinfo {author} {\bibfnamefont {K.}~\bibnamefont {Temme}}, \bibinfo
  {author} {\bibfnamefont {M.}~\bibnamefont {Takita}}, \bibinfo {author}
  {\bibfnamefont {M.}~\bibnamefont {Brink}}, \bibinfo {author} {\bibfnamefont
  {J.~M.}\ \bibnamefont {Chow}}, \ and\ \bibinfo {author} {\bibfnamefont
  {J.~M.}\ \bibnamefont {Gambetta}},\ }\href@noop {} {\bibfield  {journal}
  {\bibinfo  {journal} {Nature}\ }\textbf {\bibinfo {volume} {549}},\ \bibinfo
  {pages} {242} (\bibinfo {year} {2017})}\BibitemShut {NoStop}%
\bibitem [{\citenamefont {O’malley}\ \emph {et~al.}(2016)\citenamefont
  {O’malley}, \citenamefont {Babbush}, \citenamefont {Kivlichan},
  \citenamefont {Romero}, \citenamefont {McClean}, \citenamefont {Barends},
  \citenamefont {Kelly}, \citenamefont {Roushan}, \citenamefont {Tranter},\
  and\ \citenamefont {Ding}}]{omalley2016scalable}%
  \BibitemOpen
  \bibfield  {author} {\bibinfo {author} {\bibfnamefont {P.~J.~J.}\
  \bibnamefont {O’malley}}, \bibinfo {author} {\bibfnamefont
  {R.}~\bibnamefont {Babbush}}, \bibinfo {author} {\bibfnamefont {I.~D.}\
  \bibnamefont {Kivlichan}}, \bibinfo {author} {\bibfnamefont {J.}~\bibnamefont
  {Romero}}, \bibinfo {author} {\bibfnamefont {J.~R.}\ \bibnamefont {McClean}},
  \bibinfo {author} {\bibfnamefont {R.}~\bibnamefont {Barends}}, \bibinfo
  {author} {\bibfnamefont {J.}~\bibnamefont {Kelly}}, \bibinfo {author}
  {\bibfnamefont {P.}~\bibnamefont {Roushan}}, \bibinfo {author} {\bibfnamefont
  {A.}~\bibnamefont {Tranter}}, \ and\ \bibinfo {author} {\bibfnamefont
  {N.}~\bibnamefont {Ding}},\ }\href@noop {} {\bibfield  {journal} {\bibinfo
  {journal} {Physical Review X}\ }\textbf {\bibinfo {volume} {6}},\ \bibinfo
  {pages} {031007} (\bibinfo {year} {2016})}\BibitemShut {NoStop}%
\bibitem [{\citenamefont {Du}\ \emph {et~al.}(2010)\citenamefont {Du},
  \citenamefont {Xu}, \citenamefont {Peng}, \citenamefont {Wang}, \citenamefont
  {Wu},\ and\ \citenamefont {Lu}}]{Du2010nmr}%
  \BibitemOpen
  \bibfield  {author} {\bibinfo {author} {\bibfnamefont {J.}~\bibnamefont
  {Du}}, \bibinfo {author} {\bibfnamefont {N.}~\bibnamefont {Xu}}, \bibinfo
  {author} {\bibfnamefont {X.}~\bibnamefont {Peng}}, \bibinfo {author}
  {\bibfnamefont {P.}~\bibnamefont {Wang}}, \bibinfo {author} {\bibfnamefont
  {S.}~\bibnamefont {Wu}}, \ and\ \bibinfo {author} {\bibfnamefont
  {D.}~\bibnamefont {Lu}},\ }\href@noop {} {\bibfield  {journal} {\bibinfo
  {journal} {Physical Review Letters}\ }\textbf {\bibinfo {volume} {104}},\
  \bibinfo {pages} {030502} (\bibinfo {year} {2010})}\BibitemShut {NoStop}%
\bibitem [{\citenamefont {Hempel}\ \emph {et~al.}(2018)\citenamefont {Hempel},
  \citenamefont {Maier}, \citenamefont {Romero}, \citenamefont {McClean},
  \citenamefont {Monz}, \citenamefont {Shen}, \citenamefont {Jurcevic},
  \citenamefont {Lanyon}, \citenamefont {Love},\ and\ \citenamefont
  {Babbush}}]{hempel2018quantum}%
  \BibitemOpen
  \bibfield  {author} {\bibinfo {author} {\bibfnamefont {C.}~\bibnamefont
  {Hempel}}, \bibinfo {author} {\bibfnamefont {C.}~\bibnamefont {Maier}},
  \bibinfo {author} {\bibfnamefont {J.}~\bibnamefont {Romero}}, \bibinfo
  {author} {\bibfnamefont {J.}~\bibnamefont {McClean}}, \bibinfo {author}
  {\bibfnamefont {T.}~\bibnamefont {Monz}}, \bibinfo {author} {\bibfnamefont
  {H.}~\bibnamefont {Shen}}, \bibinfo {author} {\bibfnamefont {P.}~\bibnamefont
  {Jurcevic}}, \bibinfo {author} {\bibfnamefont {B.}~\bibnamefont {Lanyon}},
  \bibinfo {author} {\bibfnamefont {P.}~\bibnamefont {Love}}, \ and\ \bibinfo
  {author} {\bibfnamefont {R.}~\bibnamefont {Babbush}},\ }\href@noop {}
  {\bibfield  {journal} {\bibinfo  {journal} {arXiv preprint arXiv:1803.10238}\
  } (\bibinfo {year} {2018})}\BibitemShut {NoStop}%
\bibitem [{\citenamefont {Lamm}\ and\ \citenamefont
  {Lawrence}(2018)}]{lamm2018simulation}%
  \BibitemOpen
  \bibfield  {author} {\bibinfo {author} {\bibfnamefont {H.}~\bibnamefont
  {Lamm}}\ and\ \bibinfo {author} {\bibfnamefont {S.}~\bibnamefont
  {Lawrence}},\ }\href@noop {} {\bibfield  {journal} {\bibinfo  {journal}
  {Physical review letters}\ }\textbf {\bibinfo {volume} {121}},\ \bibinfo
  {pages} {170501} (\bibinfo {year} {2018})}\BibitemShut {NoStop}%
\bibitem [{\citenamefont {Wiebe}\ \emph {et~al.}(2011)\citenamefont {Wiebe},
  \citenamefont {Berry}, \citenamefont {Høyer},\ and\ \citenamefont
  {Sanders}}]{wiebe2011simulating}%
  \BibitemOpen
  \bibfield  {author} {\bibinfo {author} {\bibfnamefont {N.}~\bibnamefont
  {Wiebe}}, \bibinfo {author} {\bibfnamefont {D.~W.}\ \bibnamefont {Berry}},
  \bibinfo {author} {\bibfnamefont {P.}~\bibnamefont {Høyer}}, \ and\ \bibinfo
  {author} {\bibfnamefont {B.~C.}\ \bibnamefont {Sanders}},\ }\href@noop {}
  {\bibfield  {journal} {\bibinfo  {journal} {Journal of Physics A:
  Mathematical and Theoretical}\ }\textbf {\bibinfo {volume} {44}},\ \bibinfo
  {pages} {445308} (\bibinfo {year} {2011})}\BibitemShut {NoStop}%
\bibitem [{\citenamefont {Geim}\ and\ \citenamefont
  {Grigorieva}(2013)}]{geim2013van}%
  \BibitemOpen
  \bibfield  {author} {\bibinfo {author} {\bibfnamefont {A.~K.}\ \bibnamefont
  {Geim}}\ and\ \bibinfo {author} {\bibfnamefont {I.~V.}\ \bibnamefont
  {Grigorieva}},\ }\href@noop {} {\bibfield  {journal} {\bibinfo  {journal}
  {Nature}\ }\textbf {\bibinfo {volume} {499}},\ \bibinfo {pages} {419}
  (\bibinfo {year} {2013})}\BibitemShut {NoStop}%
\bibitem [{\citenamefont {Basov}\ \emph {et~al.}(2017)\citenamefont {Basov},
  \citenamefont {Averitt},\ and\ \citenamefont {Hsieh}}]{basov2017towards}%
  \BibitemOpen
  \bibfield  {author} {\bibinfo {author} {\bibfnamefont {D.~N.}\ \bibnamefont
  {Basov}}, \bibinfo {author} {\bibfnamefont {R.~D.}\ \bibnamefont {Averitt}},
  \ and\ \bibinfo {author} {\bibfnamefont {D.}~\bibnamefont {Hsieh}},\ }\href
  {\doibase 10.1038/NMAT5017} {\bibfield  {journal} {\bibinfo  {journal}
  {Nature Materials}\ }\textbf {\bibinfo {volume} {16}},\ \bibinfo {pages}
  {1077} (\bibinfo {year} {2017})}\BibitemShut {NoStop}%
\bibitem [{\citenamefont {Lin}\ \emph {et~al.}(2017)\citenamefont {Lin},
  \citenamefont {Kochat}, \citenamefont {Krishnamoorthy}, \citenamefont
  {Bassman}, \citenamefont {Weninger}, \citenamefont {Zheng}, \citenamefont
  {Zhang}, \citenamefont {Apte}, \citenamefont {Tiwary}, \citenamefont {Shen},
  \citenamefont {Li}, \citenamefont {Kalia}, \citenamefont {Ajayan},
  \citenamefont {Nakano}, \citenamefont {Vashishta}, \citenamefont {Shimojo},
  \citenamefont {Wang}, \citenamefont {Fritz},\ and\ \citenamefont
  {Bergmann}}]{lin2017ultrafast}%
  \BibitemOpen
  \bibfield  {author} {\bibinfo {author} {\bibfnamefont {M.~F.}\ \bibnamefont
  {Lin}}, \bibinfo {author} {\bibfnamefont {V.}~\bibnamefont {Kochat}},
  \bibinfo {author} {\bibfnamefont {A.}~\bibnamefont {Krishnamoorthy}},
  \bibinfo {author} {\bibfnamefont {L.}~\bibnamefont {Bassman}}, \bibinfo
  {author} {\bibfnamefont {C.}~\bibnamefont {Weninger}}, \bibinfo {author}
  {\bibfnamefont {Q.}~\bibnamefont {Zheng}}, \bibinfo {author} {\bibfnamefont
  {X.}~\bibnamefont {Zhang}}, \bibinfo {author} {\bibfnamefont
  {A.}~\bibnamefont {Apte}}, \bibinfo {author} {\bibfnamefont {C.~S.}\
  \bibnamefont {Tiwary}}, \bibinfo {author} {\bibfnamefont {X.~Z.}\
  \bibnamefont {Shen}}, \bibinfo {author} {\bibfnamefont {R.~K.}\ \bibnamefont
  {Li}}, \bibinfo {author} {\bibfnamefont {R.}~\bibnamefont {Kalia}}, \bibinfo
  {author} {\bibfnamefont {P.}~\bibnamefont {Ajayan}}, \bibinfo {author}
  {\bibfnamefont {A.}~\bibnamefont {Nakano}}, \bibinfo {author} {\bibfnamefont
  {P.}~\bibnamefont {Vashishta}}, \bibinfo {author} {\bibfnamefont
  {F.}~\bibnamefont {Shimojo}}, \bibinfo {author} {\bibfnamefont {X.~J.}\
  \bibnamefont {Wang}}, \bibinfo {author} {\bibfnamefont {D.~M.}\ \bibnamefont
  {Fritz}}, \ and\ \bibinfo {author} {\bibfnamefont {U.}~\bibnamefont
  {Bergmann}},\ }\href {\doibase 10.1038/s41467-017-01844-2} {\bibfield
  {journal} {\bibinfo  {journal} {Nature Communications}\ }\textbf {\bibinfo
  {volume} {8}},\ \bibinfo {pages} {1745} (\bibinfo {year} {2017})}\BibitemShut
  {NoStop}%
\bibitem [{\citenamefont {Tung}\ \emph {et~al.}(2019)\citenamefont {Tung},
  \citenamefont {Krishnamoorthy}, \citenamefont {Sadasivam}, \citenamefont
  {Zhou}, \citenamefont {Zhang}, \citenamefont {Seyler}, \citenamefont {Clark},
  \citenamefont {Mannebach}, \citenamefont {Nyby}, \citenamefont {Ernst},
  \citenamefont {Zhu}, \citenamefont {Glownia}, \citenamefont {Kozina},
  \citenamefont {Song}, \citenamefont {Nelson}, \citenamefont {Kumazoe},
  \citenamefont {Shimojo}, \citenamefont {Kalia}, \citenamefont {Vashishta},
  \citenamefont {Darancet}, \citenamefont {Heinz}, \citenamefont {Nakano},
  \citenamefont {Xu}, \citenamefont {Lindenberg},\ and\ \citenamefont
  {Wen}}]{Tung2019anisotropic}%
  \BibitemOpen
  \bibfield  {author} {\bibinfo {author} {\bibfnamefont {I.}~\bibnamefont
  {Tung}}, \bibinfo {author} {\bibfnamefont {A.}~\bibnamefont
  {Krishnamoorthy}}, \bibinfo {author} {\bibfnamefont {S.}~\bibnamefont
  {Sadasivam}}, \bibinfo {author} {\bibfnamefont {H.}~\bibnamefont {Zhou}},
  \bibinfo {author} {\bibfnamefont {Q.}~\bibnamefont {Zhang}}, \bibinfo
  {author} {\bibfnamefont {K.~L.}\ \bibnamefont {Seyler}}, \bibinfo {author}
  {\bibfnamefont {G.}~\bibnamefont {Clark}}, \bibinfo {author} {\bibfnamefont
  {E.~M.}\ \bibnamefont {Mannebach}}, \bibinfo {author} {\bibfnamefont
  {C.}~\bibnamefont {Nyby}}, \bibinfo {author} {\bibfnamefont {F.}~\bibnamefont
  {Ernst}}, \bibinfo {author} {\bibfnamefont {D.}~\bibnamefont {Zhu}}, \bibinfo
  {author} {\bibfnamefont {J.~M.}\ \bibnamefont {Glownia}}, \bibinfo {author}
  {\bibfnamefont {M.~E.}\ \bibnamefont {Kozina}}, \bibinfo {author}
  {\bibfnamefont {S.}~\bibnamefont {Song}}, \bibinfo {author} {\bibfnamefont
  {S.}~\bibnamefont {Nelson}}, \bibinfo {author} {\bibfnamefont
  {H.}~\bibnamefont {Kumazoe}}, \bibinfo {author} {\bibfnamefont
  {F.}~\bibnamefont {Shimojo}}, \bibinfo {author} {\bibfnamefont {R.~K.}\
  \bibnamefont {Kalia}}, \bibinfo {author} {\bibfnamefont {P.}~\bibnamefont
  {Vashishta}}, \bibinfo {author} {\bibfnamefont {P.}~\bibnamefont {Darancet}},
  \bibinfo {author} {\bibfnamefont {T.~F.}\ \bibnamefont {Heinz}}, \bibinfo
  {author} {\bibfnamefont {A.}~\bibnamefont {Nakano}}, \bibinfo {author}
  {\bibfnamefont {X.}~\bibnamefont {Xu}}, \bibinfo {author} {\bibfnamefont
  {A.~M.}\ \bibnamefont {Lindenberg}}, \ and\ \bibinfo {author} {\bibfnamefont
  {H.}~\bibnamefont {Wen}},\ }\href {\doibase 10.1038/s41566-019-0387-5}
  {\bibfield  {journal} {\bibinfo  {journal} {Nature Photonics}\ }\textbf
  {\bibinfo {volume} {13}},\ \bibinfo {pages} {425} (\bibinfo {year}
  {2019})}\BibitemShut {NoStop}%
\bibitem [{\citenamefont {Li}\ \emph {et~al.}(2019)\citenamefont {Li},
  \citenamefont {Qiu}, \citenamefont {Zhang}, \citenamefont {Baldini},
  \citenamefont {Lu}, \citenamefont {Rappe},\ and\ \citenamefont
  {Nelson}}]{li2019terahertz}%
  \BibitemOpen
  \bibfield  {author} {\bibinfo {author} {\bibfnamefont {X.}~\bibnamefont
  {Li}}, \bibinfo {author} {\bibfnamefont {T.}~\bibnamefont {Qiu}}, \bibinfo
  {author} {\bibfnamefont {J.~H.}\ \bibnamefont {Zhang}}, \bibinfo {author}
  {\bibfnamefont {E.}~\bibnamefont {Baldini}}, \bibinfo {author} {\bibfnamefont
  {J.}~\bibnamefont {Lu}}, \bibinfo {author} {\bibfnamefont {A.~M.}\
  \bibnamefont {Rappe}}, \ and\ \bibinfo {author} {\bibfnamefont {K.~A.}\
  \bibnamefont {Nelson}},\ }\href {\doibase 10.1126/science.aaw4913} {\bibfield
   {journal} {\bibinfo  {journal} {Science}\ }\textbf {\bibinfo {volume}
  {364}},\ \bibinfo {pages} {1079} (\bibinfo {year} {2019})}\BibitemShut
  {NoStop}%
\bibitem [{\citenamefont {Kochat}\ \emph {et~al.}(2017)\citenamefont {Kochat},
  \citenamefont {Apte}, \citenamefont {Hachtel}, \citenamefont {Kumazoe},
  \citenamefont {Krishnamoorthy}, \citenamefont {Susarla}, \citenamefont
  {Idrobo}, \citenamefont {Shimojo}, \citenamefont {Vashishta},\ and\
  \citenamefont {Kalia}}]{kochat2017re}%
  \BibitemOpen
  \bibfield  {author} {\bibinfo {author} {\bibfnamefont {V.}~\bibnamefont
  {Kochat}}, \bibinfo {author} {\bibfnamefont {A.}~\bibnamefont {Apte}},
  \bibinfo {author} {\bibfnamefont {J.~A.}\ \bibnamefont {Hachtel}}, \bibinfo
  {author} {\bibfnamefont {H.}~\bibnamefont {Kumazoe}}, \bibinfo {author}
  {\bibfnamefont {A.}~\bibnamefont {Krishnamoorthy}}, \bibinfo {author}
  {\bibfnamefont {S.}~\bibnamefont {Susarla}}, \bibinfo {author} {\bibfnamefont
  {J.~C.}\ \bibnamefont {Idrobo}}, \bibinfo {author} {\bibfnamefont
  {F.}~\bibnamefont {Shimojo}}, \bibinfo {author} {\bibfnamefont
  {P.}~\bibnamefont {Vashishta}}, \ and\ \bibinfo {author} {\bibfnamefont
  {R.}~\bibnamefont {Kalia}},\ }\href@noop {} {\bibfield  {journal} {\bibinfo
  {journal} {Advanced Materials}\ }\textbf {\bibinfo {volume} {29}},\ \bibinfo
  {pages} {1703754} (\bibinfo {year} {2017})}\BibitemShut {NoStop}%
\bibitem [{\citenamefont {Shin}\ \emph {et~al.}(2018)\citenamefont {Shin},
  \citenamefont {Hübener}, \citenamefont {De~Giovannini}, \citenamefont {Jin},
  \citenamefont {Rubio},\ and\ \citenamefont {Park}}]{shin2018phonon}%
  \BibitemOpen
  \bibfield  {author} {\bibinfo {author} {\bibfnamefont {D.}~\bibnamefont
  {Shin}}, \bibinfo {author} {\bibfnamefont {H.}~\bibnamefont {Hübener}},
  \bibinfo {author} {\bibfnamefont {U.}~\bibnamefont {De~Giovannini}}, \bibinfo
  {author} {\bibfnamefont {H.}~\bibnamefont {Jin}}, \bibinfo {author}
  {\bibfnamefont {A.}~\bibnamefont {Rubio}}, \ and\ \bibinfo {author}
  {\bibfnamefont {N.}~\bibnamefont {Park}},\ }\href@noop {} {\bibfield
  {journal} {\bibinfo  {journal} {Nature Communications}\ }\textbf {\bibinfo
  {volume} {9}},\ \bibinfo {pages} {638} (\bibinfo {year} {2018})}\BibitemShut
  {NoStop}%
\bibitem [{\citenamefont {Bassman}\ \emph {et~al.}(2018)\citenamefont
  {Bassman}, \citenamefont {Krishnamoorthy}, \citenamefont {Kumazoe},
  \citenamefont {Misawa}, \citenamefont {Shimojo}, \citenamefont {Kalia},
  \citenamefont {Nakano},\ and\ \citenamefont
  {Vashishta}}]{bassman2018electronic}%
  \BibitemOpen
  \bibfield  {author} {\bibinfo {author} {\bibfnamefont {L.}~\bibnamefont
  {Bassman}}, \bibinfo {author} {\bibfnamefont {A.}~\bibnamefont
  {Krishnamoorthy}}, \bibinfo {author} {\bibfnamefont {H.}~\bibnamefont
  {Kumazoe}}, \bibinfo {author} {\bibfnamefont {M.}~\bibnamefont {Misawa}},
  \bibinfo {author} {\bibfnamefont {F.}~\bibnamefont {Shimojo}}, \bibinfo
  {author} {\bibfnamefont {R.~K.}\ \bibnamefont {Kalia}}, \bibinfo {author}
  {\bibfnamefont {A.}~\bibnamefont {Nakano}}, \ and\ \bibinfo {author}
  {\bibfnamefont {P.}~\bibnamefont {Vashishta}},\ }\href@noop {} {\bibfield
  {journal} {\bibinfo  {journal} {Nano letters}\ }\textbf {\bibinfo {volume}
  {18}},\ \bibinfo {pages} {4653} (\bibinfo {year} {2018})}\BibitemShut
  {NoStop}%
\bibitem [{\citenamefont {Poulin}\ \emph {et~al.}(2011)\citenamefont {Poulin},
  \citenamefont {Qarry}, \citenamefont {Somma},\ and\ \citenamefont
  {Verstraete}}]{poulin2011quantum}%
  \BibitemOpen
  \bibfield  {author} {\bibinfo {author} {\bibfnamefont {D.}~\bibnamefont
  {Poulin}}, \bibinfo {author} {\bibfnamefont {A.}~\bibnamefont {Qarry}},
  \bibinfo {author} {\bibfnamefont {R.}~\bibnamefont {Somma}}, \ and\ \bibinfo
  {author} {\bibfnamefont {F.}~\bibnamefont {Verstraete}},\ }\href@noop {}
  {\bibfield  {journal} {\bibinfo  {journal} {Physical Review Letters}\
  }\textbf {\bibinfo {volume} {106}},\ \bibinfo {pages} {170501} (\bibinfo
  {year} {2011})}\BibitemShut {NoStop}%
\bibitem [{\citenamefont {Trotter}(1959)}]{trotter1959on}%
  \BibitemOpen
  \bibfield  {author} {\bibinfo {author} {\bibfnamefont {H.~F.}\ \bibnamefont
  {Trotter}},\ }\href@noop {} {\bibfield  {journal} {\bibinfo  {journal}
  {Proceedings of the American Mathematical Society}\ }\textbf {\bibinfo
  {volume} {10}},\ \bibinfo {pages} {545} (\bibinfo {year} {1959})}\BibitemShut
  {NoStop}%
\bibitem [{\citenamefont {Pfeuty}(1970)}]{pfeuty1970one}%
  \BibitemOpen
  \bibfield  {author} {\bibinfo {author} {\bibfnamefont {P.}~\bibnamefont
  {Pfeuty}},\ }\href@noop {} {\bibfield  {journal} {\bibinfo  {journal} {ANNALS
  of Physics}\ }\textbf {\bibinfo {volume} {57}},\ \bibinfo {pages} {79}
  (\bibinfo {year} {1970})}\BibitemShut {NoStop}%
\bibitem [{\citenamefont {Lieb}\ \emph {et~al.}(1961)\citenamefont {Lieb},
  \citenamefont {Schultz},\ and\ \citenamefont {Mattis}}]{lieb1961two}%
  \BibitemOpen
  \bibfield  {author} {\bibinfo {author} {\bibfnamefont {E.}~\bibnamefont
  {Lieb}}, \bibinfo {author} {\bibfnamefont {T.}~\bibnamefont {Schultz}}, \
  and\ \bibinfo {author} {\bibfnamefont {D.}~\bibnamefont {Mattis}},\
  }\href@noop {} {\bibfield  {journal} {\bibinfo  {journal} {Annals of
  Physics}\ }\textbf {\bibinfo {volume} {16}},\ \bibinfo {pages} {407}
  (\bibinfo {year} {1961})}\BibitemShut {NoStop}%
\bibitem [{\citenamefont {Verstraete}\ \emph {et~al.}(2009)\citenamefont
  {Verstraete}, \citenamefont {Cirac},\ and\ \citenamefont
  {Latorre}}]{verstraete2009quantum}%
  \BibitemOpen
  \bibfield  {author} {\bibinfo {author} {\bibfnamefont {F.}~\bibnamefont
  {Verstraete}}, \bibinfo {author} {\bibfnamefont {J.~I.}\ \bibnamefont
  {Cirac}}, \ and\ \bibinfo {author} {\bibfnamefont {J.~I.}\ \bibnamefont
  {Latorre}},\ }\href@noop {} {\bibfield  {journal} {\bibinfo  {journal}
  {Physical Review A}\ }\textbf {\bibinfo {volume} {79}},\ \bibinfo {pages}
  {032316} (\bibinfo {year} {2009})}\BibitemShut {NoStop}%
\bibitem [{\citenamefont {Cervera-Lierta}(2018)}]{cervera2018exact}%
  \BibitemOpen
  \bibfield  {author} {\bibinfo {author} {\bibfnamefont {A.}~\bibnamefont
  {Cervera-Lierta}},\ }\href@noop {} {\bibfield  {journal} {\bibinfo  {journal}
  {Quantum}\ }\textbf {\bibinfo {volume} {2}},\ \bibinfo {pages} {114}
  (\bibinfo {year} {2018})}\BibitemShut {NoStop}%
\bibitem [{\citenamefont {Mosca}\ and\ \citenamefont
  {Kaye}(2001)}]{mosca2001quantum}%
  \BibitemOpen
  \bibfield  {author} {\bibinfo {author} {\bibfnamefont {M.}~\bibnamefont
  {Mosca}}\ and\ \bibinfo {author} {\bibfnamefont {P.}~\bibnamefont {Kaye}},\
  }in\ \href@noop {} {\emph {\bibinfo {booktitle} {International Conference on
  Quantum Information}}}\ (\bibinfo {organization} {Optical Society of
  America},\ \bibinfo {year} {2001})\ p.\ \bibinfo {pages} {PB28}\BibitemShut
  {NoStop}%
\bibitem [{\citenamefont {Soklakov}\ and\ \citenamefont
  {Schack}(2006)}]{soklakov2006efficient}%
  \BibitemOpen
  \bibfield  {author} {\bibinfo {author} {\bibfnamefont {A.~N.}\ \bibnamefont
  {Soklakov}}\ and\ \bibinfo {author} {\bibfnamefont {R.}~\bibnamefont
  {Schack}},\ }\href@noop {} {\bibfield  {journal} {\bibinfo  {journal}
  {Physical Review A}\ }\textbf {\bibinfo {volume} {73}},\ \bibinfo {pages}
  {012307} (\bibinfo {year} {2006})}\BibitemShut {NoStop}%
\bibitem [{\citenamefont {Kitaev}\ and\ \citenamefont
  {Webb}(2008)}]{kitaev2008wavefunction}%
  \BibitemOpen
  \bibfield  {author} {\bibinfo {author} {\bibfnamefont {A.}~\bibnamefont
  {Kitaev}}\ and\ \bibinfo {author} {\bibfnamefont {W.~A.}\ \bibnamefont
  {Webb}},\ }\href@noop {} {\bibfield  {journal} {\bibinfo  {journal} {arXiv
  preprint arXiv:0801.0342}\ } (\bibinfo {year} {2008})}\BibitemShut {NoStop}%
\bibitem [{\citenamefont {Wang}\ \emph {et~al.}(2009)\citenamefont {Wang},
  \citenamefont {Ashhab},\ and\ \citenamefont {Nori}}]{wang2009efficient}%
  \BibitemOpen
  \bibfield  {author} {\bibinfo {author} {\bibfnamefont {H.}~\bibnamefont
  {Wang}}, \bibinfo {author} {\bibfnamefont {S.}~\bibnamefont {Ashhab}}, \ and\
  \bibinfo {author} {\bibfnamefont {F.}~\bibnamefont {Nori}},\ }\href@noop {}
  {\bibfield  {journal} {\bibinfo  {journal} {Physical Review A}\ }\textbf
  {\bibinfo {volume} {79}},\ \bibinfo {pages} {042335} (\bibinfo {year}
  {2009})}\BibitemShut {NoStop}%
\bibitem [{\citenamefont {Veis}\ and\ \citenamefont
  {Pittner}(2014)}]{Veis2014adiabatic}%
  \BibitemOpen
  \bibfield  {author} {\bibinfo {author} {\bibfnamefont {L.}~\bibnamefont
  {Veis}}\ and\ \bibinfo {author} {\bibfnamefont {J.}~\bibnamefont {Pittner}},\
  }\href@noop {} {\bibfield  {journal} {\bibinfo  {journal} {Journal of
  Chemical Physics}\ }\textbf {\bibinfo {volume} {140}},\ \bibinfo {pages}
  {214111} (\bibinfo {year} {2014})}\BibitemShut {NoStop}%
\bibitem [{\citenamefont {Tubman}\ \emph {et~al.}(2018)\citenamefont {Tubman},
  \citenamefont {Mejuto-Zaera}, \citenamefont {Epstein}, \citenamefont {Hait},
  \citenamefont {Levine}, \citenamefont {Huggins}, \citenamefont {Jiang},
  \citenamefont {McClean}, \citenamefont {Babbush},\ and\ \citenamefont
  {Head-Gordon}}]{tubman2018postponing}%
  \BibitemOpen
  \bibfield  {author} {\bibinfo {author} {\bibfnamefont {N.~M.}\ \bibnamefont
  {Tubman}}, \bibinfo {author} {\bibfnamefont {C.}~\bibnamefont
  {Mejuto-Zaera}}, \bibinfo {author} {\bibfnamefont {J.~M.}\ \bibnamefont
  {Epstein}}, \bibinfo {author} {\bibfnamefont {D.}~\bibnamefont {Hait}},
  \bibinfo {author} {\bibfnamefont {D.~S.}\ \bibnamefont {Levine}}, \bibinfo
  {author} {\bibfnamefont {W.}~\bibnamefont {Huggins}}, \bibinfo {author}
  {\bibfnamefont {Z.}~\bibnamefont {Jiang}}, \bibinfo {author} {\bibfnamefont
  {J.~R.}\ \bibnamefont {McClean}}, \bibinfo {author} {\bibfnamefont
  {R.}~\bibnamefont {Babbush}}, \ and\ \bibinfo {author} {\bibfnamefont
  {M.}~\bibnamefont {Head-Gordon}},\ }\href@noop {} {\bibfield  {journal}
  {\bibinfo  {journal} {arXiv preprint arXiv:1809.05523}\ } (\bibinfo {year}
  {2018})}\BibitemShut {NoStop}%
\bibitem [{\citenamefont {Holthaus}(1992)}]{holthaus1992collapse}%
  \BibitemOpen
  \bibfield  {author} {\bibinfo {author} {\bibfnamefont {M.}~\bibnamefont
  {Holthaus}},\ }\href {\doibase 10.1103/PhysRevLett.69.351} {\bibfield
  {journal} {\bibinfo  {journal} {Physical Review Letters}\ }\textbf {\bibinfo
  {volume} {69}},\ \bibinfo {pages} {351} (\bibinfo {year} {1992})}\BibitemShut
  {NoStop}%
\bibitem [{\citenamefont {Floquet}(1883)}]{floquet1883equations}%
  \BibitemOpen
  \bibfield  {author} {\bibinfo {author} {\bibfnamefont {G.}~\bibnamefont
  {Floquet}},\ }in\ \href@noop {} {\emph {\bibinfo {booktitle} {Annales
  scientifiques de l'{\'E}cole normale sup{\'e}rieure}}},\ Vol.~\bibinfo
  {volume} {12}\ (\bibinfo {year} {1883})\ pp.\ \bibinfo {pages}
  {47--88}\BibitemShut {NoStop}%
\bibitem [{\citenamefont {Bukov}\ \emph {et~al.}(2015)\citenamefont {Bukov},
  \citenamefont {D'Alessio},\ and\ \citenamefont
  {Polkovnikov}}]{bukov2015universal}%
  \BibitemOpen
  \bibfield  {author} {\bibinfo {author} {\bibfnamefont {M.}~\bibnamefont
  {Bukov}}, \bibinfo {author} {\bibfnamefont {L.}~\bibnamefont {D'Alessio}}, \
  and\ \bibinfo {author} {\bibfnamefont {A.}~\bibnamefont {Polkovnikov}},\
  }\href@noop {} {\bibfield  {journal} {\bibinfo  {journal} {Advances in
  Physics}\ }\textbf {\bibinfo {volume} {64}},\ \bibinfo {pages} {139}
  (\bibinfo {year} {2015})}\BibitemShut {NoStop}%
\bibitem [{\citenamefont {Venuti}\ \emph {et~al.}(2017)\citenamefont {Venuti},
  \citenamefont {Ma}, \citenamefont {Saleur},\ and\ \citenamefont
  {Haas}}]{venuti2017topological}%
  \BibitemOpen
  \bibfield  {author} {\bibinfo {author} {\bibfnamefont {L.~C.}\ \bibnamefont
  {Venuti}}, \bibinfo {author} {\bibfnamefont {Z.~Z.}\ \bibnamefont {Ma}},
  \bibinfo {author} {\bibfnamefont {H.}~\bibnamefont {Saleur}}, \ and\ \bibinfo
  {author} {\bibfnamefont {S.}~\bibnamefont {Haas}},\ }\href {\doibase
  10.1103/PhysRevA.96.053858} {\bibfield  {journal} {\bibinfo  {journal}
  {Physical Review A}\ }\textbf {\bibinfo {volume} {96}},\ \bibinfo {pages}
  {053858} (\bibinfo {year} {2017})}\BibitemShut {NoStop}%
\bibitem [{\citenamefont {Valenzuela}\ and\ \citenamefont
  {Roche}(2019)}]{valenzuela2019phase}%
  \BibitemOpen
  \bibfield  {author} {\bibinfo {author} {\bibfnamefont {S.~O.}\ \bibnamefont
  {Valenzuela}}\ and\ \bibinfo {author} {\bibfnamefont {S.}~\bibnamefont
  {Roche}},\ }\href@noop {} {\bibfield  {journal} {\bibinfo  {journal} {Nature
  Nanotechnology}\ }\textbf {\bibinfo {volume} {14}},\ \bibinfo {pages} {1088}
  (\bibinfo {year} {2019})}\BibitemShut {NoStop}%
\end{thebibliography}%

\end{document}